\documentclass{acmsiggraph}               % final
\usepackage{mathptmx}
\usepackage{graphicx}
\usepackage{times}
\usepackage{subcaption}
\usepackage[hidelinks]{hyperref}
\usepackage{amsmath}
\usepackage{flushend}
\usepackage{array}

\onlineid{0}

\acmformat{print}

\title{Kernel Projection of Latent Structures Regression for Facial Animation Retargeting}

\author{Christos Ouzounis$^1$\thanks{christos.ouzounis@hs-owl.de} \qquad  Alex Kilias$^2$\thanks{alexk@kent.ac.uk}  \qquad Christos Mousas$^3$\thanks{christos@cs.siu.edu}\\
\\
$^1$Dept. of Media Production, Ostwestfalen-Lippe University of Applied Sciences, 32657 Lemgo, Germany\\
$^2$School of Engineering and Digital Arts, University of Kent, Canterbury CT2 7NT, UK\\
$^3$Dept. of Computer Science, Southern Illinois University, Carbondale, IL 62901, USA\\
}


\begin{document}

\teaser{
\includegraphics[height=0.25\textwidth]{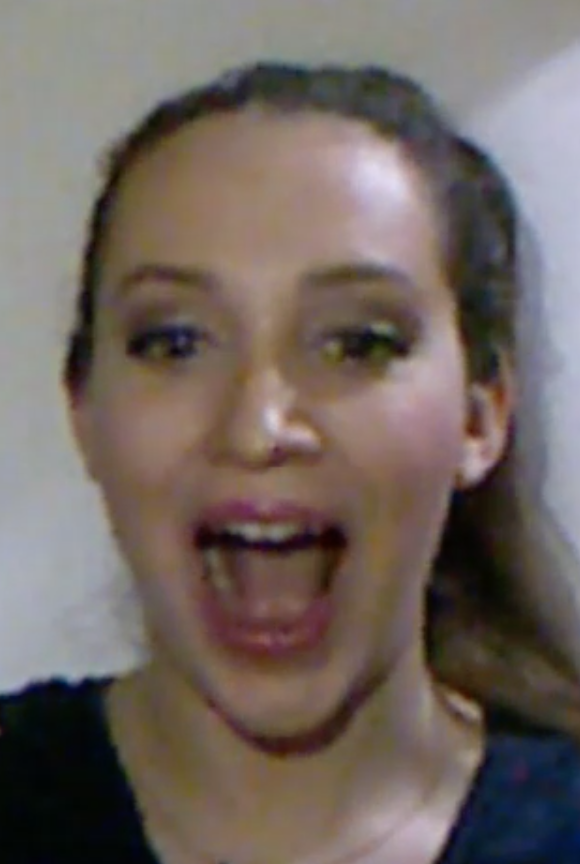}
      \includegraphics[height=0.25\textwidth]{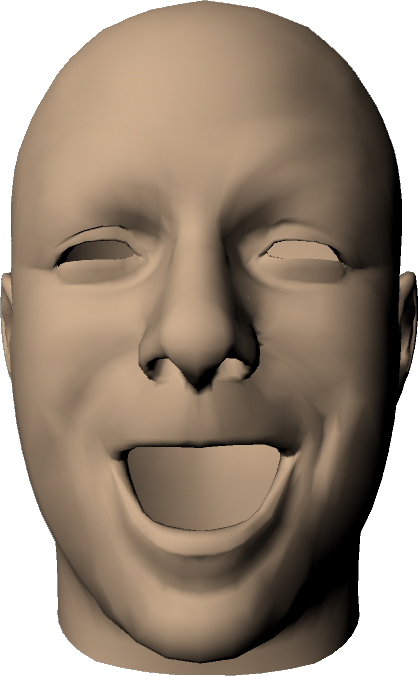}
      \includegraphics[height=0.25\textwidth]{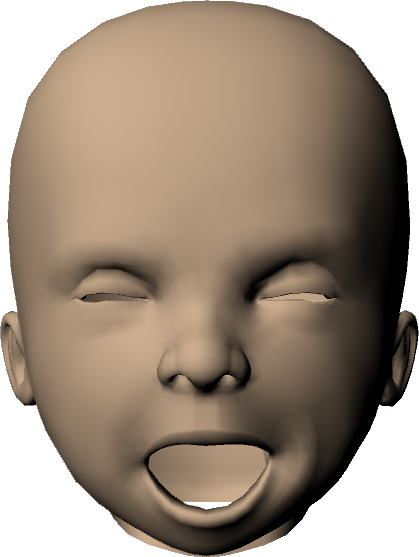}
  \centering
   \caption{The methodology that is presented provides the ability to efficiently transfer facial animations to characters with different morphological variations.}
 \label{fig1}
 }

%% The ``\maketitle'' command must be the first command after the
%% ``\begin{document}'' command. It prepares and prints the title block.

\maketitle

%% Abstract section.

\begin{abstract}
Inspired by kernel methods that have been used extensively in achieving efficient facial animation retargeting, this paper presents a solution to retargeting facial animation in virtual character's face model based on the kernel projection of latent structure (KPLS) regression between semantically similar facial expressions. Specifically, a given number of corresponding semantically similar facial expressions are projected into the latent space. By using the Nonlinear Iterative Partial Least Square method, decomposition of the latent variables is achieved. Finally, the KPLS is achieved by solving a kernalized version of the eigenvalue problem. By evaluating our methodology with other kernel-based solutions, the efficiency of the presented methodology in transferring facial animation to face models with different morphological variations is demonstrated. 

\end{abstract}

%% ACM Computing Review (CR) categories. 
%% See <http://www.acm.org/class/1998/> for details.
%% The ``\CRcat'' command takes four arguments.

\begin{CRcatlist}
  \CRcat{I.3.7}{Computer Graphics}{Three-Dimensional Graphics and Realism}{Animation}
\end{CRcatlist}

%% The ``\keywordlist'' command prints out the keywords.
\keywordlist

\section{Introduction}
\label{sec1}

%% The ``\copyrightspace'' command must be the first command after the 
%% start of the first section of the body of your paper. It ensures the
%% copyright space is left at the bottom of the first column on the first
%% page of your paper.

\copyrightspace

Expressive facial motion is always important, especially in films and video games in which virtual characters evolve. Generally, human characteristics and emotions for at least a number of basic emotions can be sufficiently recognized by observing movements and static postures of the whole human body \cite{ref37}\cite{ref38}. However, facial expressions present an additional factor which demands a correct emotion recognition process and a greater number of emotions \cite{ref29}\cite{ref47} than full-body motion. Moreover, facial expressions can also represent significant features of human communication \cite{ref41}\cite{ref50}. This principle can also be observed in applications in which virtual characters evolve, such as in films and videogames \cite{ref32}\cite{ref33}\cite{ref35}. Moreover, a variety of perceptual studies \cite{ref53}\cite{ref54} indicate that the meaning of a character?s motion is enhanced when adding finger \cite{ref55}\cite{ref56}\cite{ref57} and facial expressions to the full-body motion of a virtual character. Therefore, it can be stated that facial animation, when applied to virtual characters, enhances their appeal, realism and credibility.

Designing highly realistic facial animations for virtual characters can be described as a time-consuming and complex process that requires talent and specialized skills. Today, with the rapid development of motion capture systems, one is able to instantly capture an actor's performance of the required motions. The facial motion capture process decreases the time that is required for the manually facial animation process, which is generally based on key-frame techniques.

Having captured the required facial motion sequence, this motion data can be applied to a virtual face model. This process is known as animation retargeting. Retargeting in facial animation requires the mapping of the captured motion of an actor in relation to the actual expressions that a virtual character can reproduce. However, the process presents two basic disadvantages. Firstly, since humans can perform a vast number of different facial expressions, it is very difficult to design all of the expressions that are required for a virtual character. Hence, a reduced number of facial morphs, the so-called blendshapes, are used for as many of the human facial expressions as possible. Because a character may require nearly one hundred facial morphs and, more than ten characters may appear in a film, blendshape transfer techniques \cite{ref2}\cite{ref3} have been developed to cope with the automatic generation of the facial morphs required for multiple characters. A second disadvantage arises in facial motion capture because human actors normally are used to performing the required motions. This occurs when dealing with characters whose facial morphological characteristics (e.g., non-human characters and consequently their face models) are not similar to those of the actor's face. Thus, careful consideration of the actor's facial expressions and the expressions that a character's face model can produce is required for the aforementioned mapping process.

In an example-based facial animation retargeting process, it is necessary to establish correspondences between the facial expressions of different characters. For that reason, a high-level of semantic knowledge of the expression spaces between the actor and the face model is required. Generally, low-level correspondence-related automatic retargeting methods fail. The reason is that the correspondences are not sufficient or built properly. Hence, methodologies that provide sets of explicit correspondence points between these dissimilar motion spaces have been used extensively to solve such problems. A typical example of such a correspondence could be a smile expression performed by an actor and a semantically similar smile of the virtual character that a modeler has modeled or sculptured. Having the semantic correspondence, the retargeting process is generally assigned to a scattered point approximation problem \cite{ref9}.

In the presented methodology, the first step is to define correspondences between input captured animation and its associated source facial expressions. The correspondences are constructed by using a number of feature points (position of vertices) that are provided by the face models and projected into the latent space. Then, the retargeting process is assigned to a kernel method that is based on the projection of latent structures between examples of semantically similar expression pairs. The advantage of such a methodology is mainly its ability to maintain the correspondence between data samples, while asking the system to align the source and target facial expressions in the latent space. This process enables the mapped expression pairs to cope effectively with a facial motion sequence, while maintaining the morphological different expressions of the target face model. By evaluating the methodology based on different kernel-based method, one can see that the facial expressions that have been captured by an actor have been effectively retargeted onto different face models. A simple example of the presented methodology by which a facial expression was retargeted onto characters with different morphological variations appears in Figure \ref{fig1}. The remainder of this paper is organized in the following manner. Section \ref{sec2} provides related work on facial animation retargeting. Section \ref{sec3} describes the proposed facial animation retargeting method. Section \ref{sec4} presents the evaluations that were conducted in conjunction with the associated results. Finally, conclusions are drawn and potential future work is discussed in Section \ref{sec5}.

%-------------------------------------------------------------------------
\section{Related Work}
\label{sec2}

During the past years, a number of papers have been published on facial animation. A comprehensive background of the techniques that have been used to animate face models can be found in \cite{ref4}\cite{ref5}. In the facial animation pipeline, there are generally a number of experts in different fields who must be involved in order to produce the final animation of a virtual character. In addition to other experts, a modeler, a rigger and an animator are always required to model, rig and animate the character. However, a variety of methodologies on ways to automate the content creation and animation pipeline have been proposed during recent years. Specifically, instead of modeling the face model of a character by hand, one can simply capture and reconstruct the mesh \cite{ref42}\cite{ref43} using 3D scanners or RGBD sensors. In addition, rather than rigging a character by hand, one can rig the character automatically by example-based rigging techniques \cite{ref44}\cite{ref45}. Finally, instead of synthesizing the desired motion of the virtual characters using time-consuming keyframe techniques, it is possible to record the required motion directly using a motion capture system \cite{ref16}\cite{ref26}.

After the aforementioned steps, a key role in the facial animation pipeline is the animation retargeting process \cite{ref9}. Specifically, in facial animation retargeting it is necessary to map into a virtual face model the captured facial expressions of an actor. Among other methodologies, the one that Williams proposed \cite{ref8} can be described as the flagship of facial animation retargeting. Generally, the aforementioned solution usually is characterized by its simplicity. Therefore, it is ideal for mapping the facial expressions that a performer has captured in a face model that has similar morphological characteristics.

Following the introductory facial animation retargeting methodology of Williams \cite{ref8}, a variety of other methodologies have been proposed in recent years. Generally, these methodologies are based on correspondence between captured markers and targeted facial expressions \cite{ref10}\cite{ref12}. The basic disadvantage of such solutions is the need for the actor and the virtual face model to have similar facial geometry. Dense correspondence between source and target face models has also been examined \cite{ref13}\cite{ref14}. In these approaches, vertex or triangle motion transfer is used to retarget the facial expressions. However, due to the linearity of the blendshape model, the reproduction of subtle non-linear motion is mainly a disadvantage for the solutions that were mentioned previously. It should be noted that non-linear model such as the kernel canonical correlation analysis (kCCA) has been used in facial deformation transfer by Feng et al. \cite{ref52}.

A variety of facial motion capture and retargeting systems \cite{ref12}\cite{ref15}\cite{ref16} use a representation of blendshapes. This is known as Facial Action Coding System (FACS) \cite{ref17}. Generally, the previously mentioned methodologies establish mappings between examples of facial expressions of a user and examples of facial expressions represented by blendshapes that are then used to control a face model. Improvements in these methodologies have been proposed in \cite{ref25}\cite{ref26}. Specifically, in these methods, dynamic expression models with online tracking demonstrate impressive tracking and animation results for an arbitrary number of users without any training or calibration. However, the initial application of such systems was not one for facial animation retargeting.

The presented methodology is quite similar to the previous example-based methods, such as \cite{ref18}\cite{ref19}\cite{ref20}\cite{ref21} and \cite{ref22}, which do not require that the meshes of a source and the target face models be geometrically similar. However, in the presented method, the correspondence between facial expressions and existing blendshapes are defined by assigning the retargeting process to a kernalized method for the projections of latent structures of facial expressions. By aligning the latent structure between the source and the target facial expressions, the presented method maintains the structure of the feature points that are used. The advantage of such a representation is its runtime efficiency and, consequently, it's easy applicability to facial animation retargeting.

%-------------------------------------------------------------------------
\section{Methodology}
\label{sec3}

This section presents the proposed methodology that was developed to retarget facial animation into virtual character's face models. 

\subsection{Representation of Expressions}
\label{sec31}

\begin{figure}[htb]
  \centering
      \includegraphics[width=1\columnwidth]{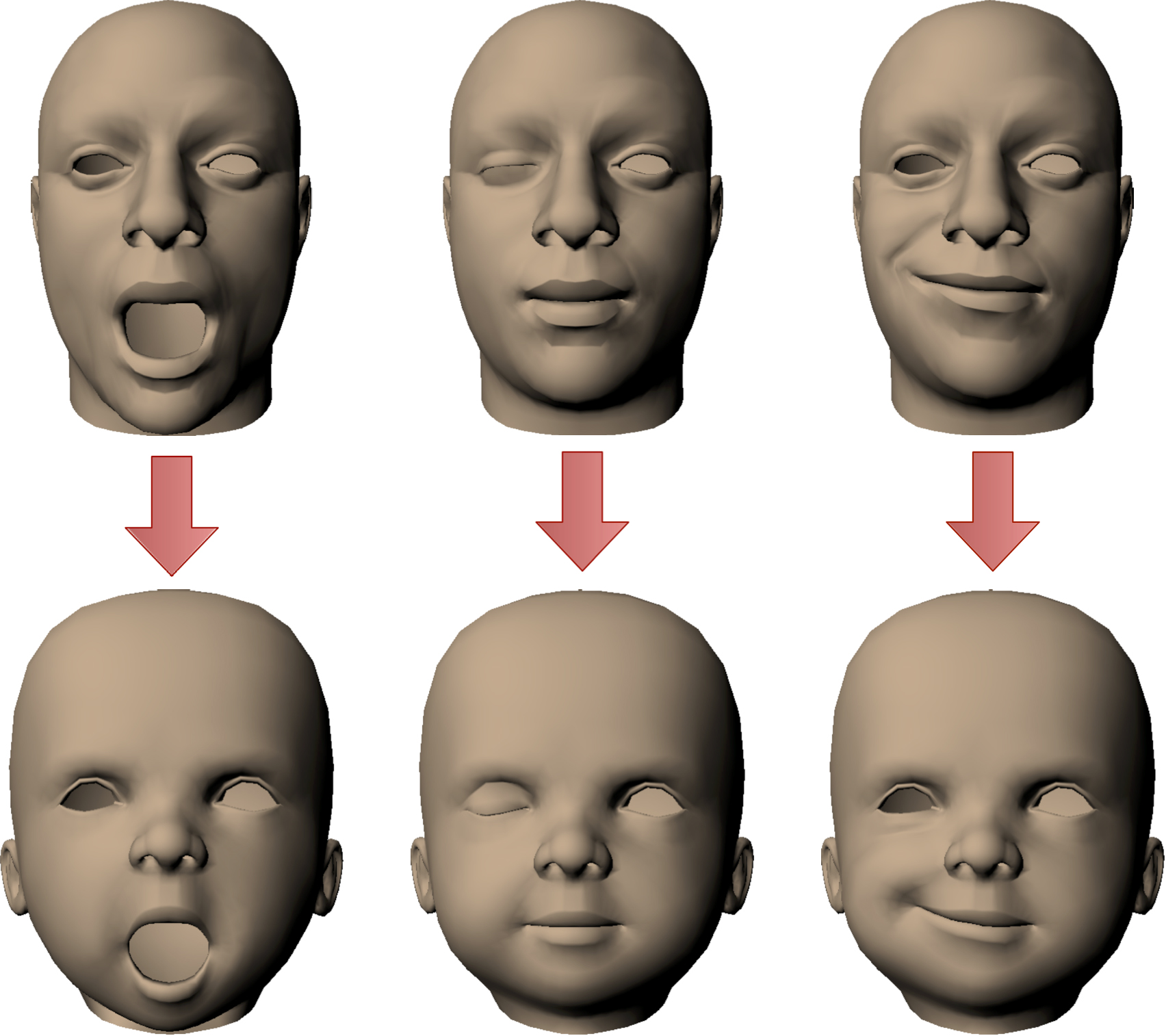}
  \caption{Examples of semantically similar facial expression pairs that are drawn from two different face models.}
   \label{fig2}
\end{figure}

What we need to consider first in the animation retargeting process is a number of source S$=\{s_1,...,s_N\}$ and target $T=\{t_1,...,t_N\}$ expression pairs that can be described as semantically similar (see Figure \ref{fig2}). These expressions can be manually modeled, sculptured or generated by the use of example-based blendshape transfer techniques \cite{ref2}\cite{ref3}. A number of feature points are selected manually in both the source and reference expressions. Hence, each $s_i$ and $t_i$ facial expression can be represented as a number of feature points $X=\{x_1,...,x_X\}$ and $Y=\{y_1,...,y_L\}$ for the source and the target face model, respectively. In the proposed methodology, a different number of feature points are used for each face model as presented in this paper (see Table \ref{tab1} and Figure \ref{fig4}).

\subsection{Animation Retargeting}
\label{sec32}
The presented facial animation retargeting method is based on the KPLS regression technique. Here, it is presented in the way in which the semantically similar facial expressions are mapped. Generally, the PLS regression assumes that the input and output datasets are related to a similar set of latent variables. The projection of latent variables of the $S$ and $T$ datasets are represented by:
\begin{equation}
S \cong G \times P^g
\label{eq1}
\end{equation}
\begin{equation}
T \cong U \times Q^g
\label{eq2}
\end{equation}
where the dimension of $S$ and $T$ is $n \times p$, the dimensionality of $P$ is $n_s \times p$, and the dimensionality of $Q$ is $n_t \times p$.

To compute the decomposition of Equations \ref{eq1} and \ref{eq2}, we use the Nonlinear Iterative Partial Least Square Method (NIPALS) \cite{ref1}, which performs a number of $p$ iterations that are divided into three steps. In the first step, it is necessary to find the linear combinations of columns of $S$ and $T$ that are maximally correlated. This is achieved by minimizing the following function:
\begin{equation}
\arg\max_{\|w\|=\|c\|=1} [cov(Sw,Tc)]^2 
\label{eq3}
\end{equation}
Then, in the second step, the latent vectors are computed by:
\begin{equation}
g = S \times r
\label{eq4}
\end{equation}
\begin{equation}
u = T \times z
\label{eq5}
\end{equation}
Finally, in the third, and final step, $S$ and $T$ are deflated by their rank-$1$ approximations based on $g$'s direction $d=g/\|g\|$ as:
\begin{equation}
S = S - dd^g \times S
\label{eq6}
\end{equation}
\begin{equation}
T = T - dd^g \times T
\label{eq7}
\end{equation}
By concatenating all $p$ vectors $g$ and $u$ into matrices of $G$ and $U$, the output prediction, $t^*$, of an input facial expression, $s^*$, is estimated by:
\begin{equation}
t^*=T^g G(U^g SS^g G)^{-1} U^g Ss^*
\label{eq8}
\end{equation}
However, the KPLS is achieved by solving a kernalized version of the eigenvalue problem in Equation \ref{eq3} and updating a Gram matrix instead of $S$ as: 
\begin{equation}
K = K - dd^g K - K dd^g + dd^g K dd^g
\label{eq9}
\end{equation}
Based on the kernalized representation of PLS, the final output prediction (retargeted motion), $t^*$, of the input facial expression, $s^*$, is achieved by computing:
\begin{equation}
t^* = T^g G(U^g K_a (S)G)^{-1} U^g K_a (s^*)
\label{eq10}
\end{equation}

\begin{figure}[!h]
  \centering
      \includegraphics[height=0.47\columnwidth]{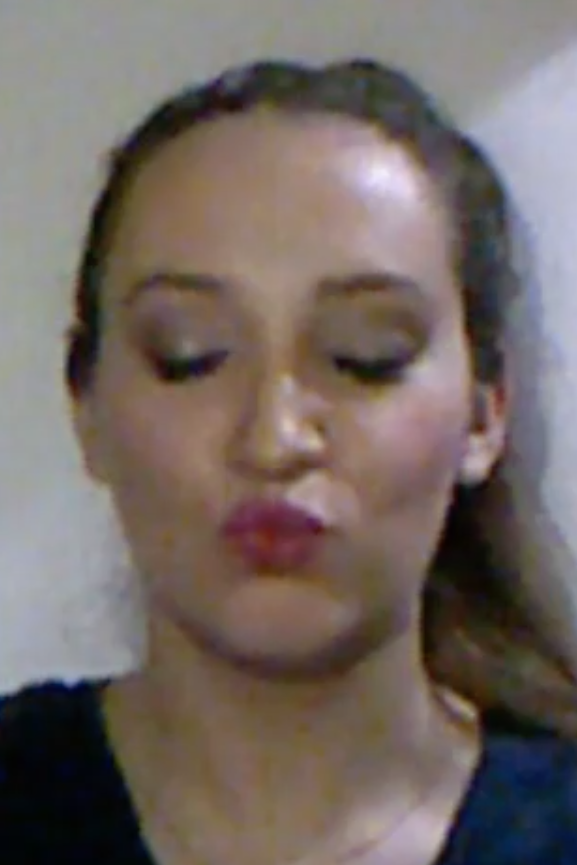}
      \includegraphics[height=0.47\columnwidth]{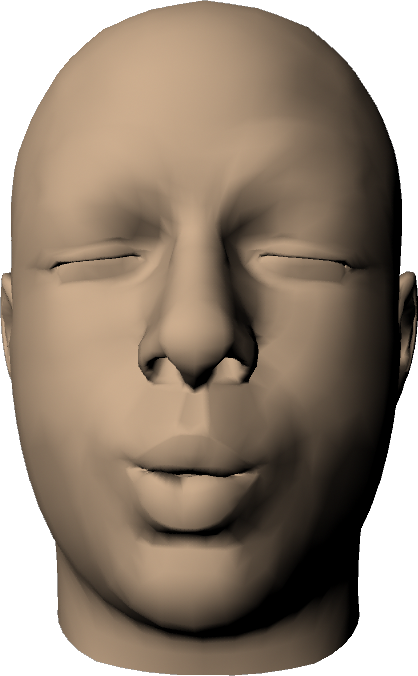}
      \includegraphics[height=0.47\columnwidth]{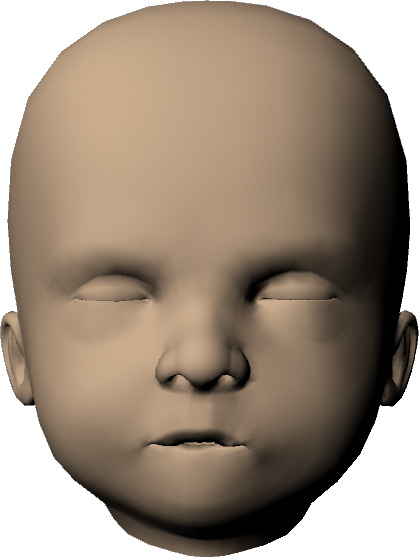}
      \includegraphics[height=0.47\columnwidth]{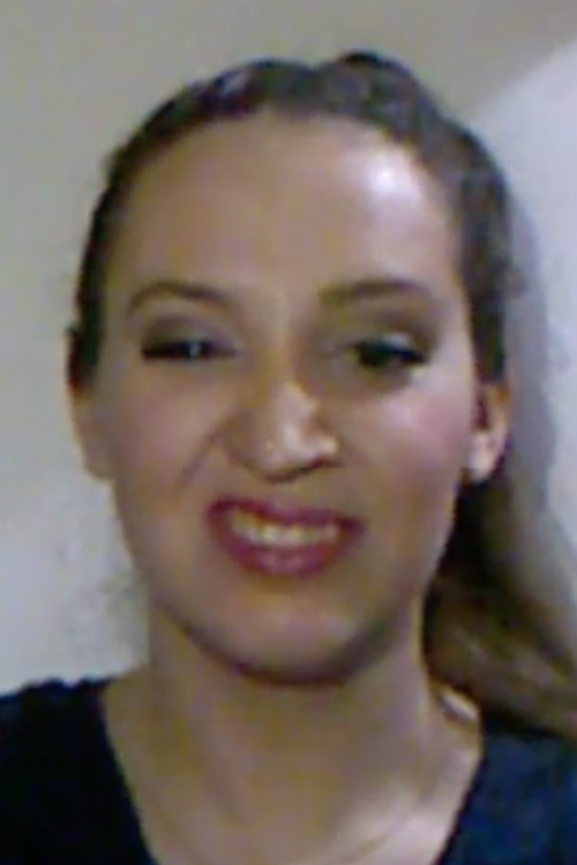}
      \includegraphics[height=0.47\columnwidth]{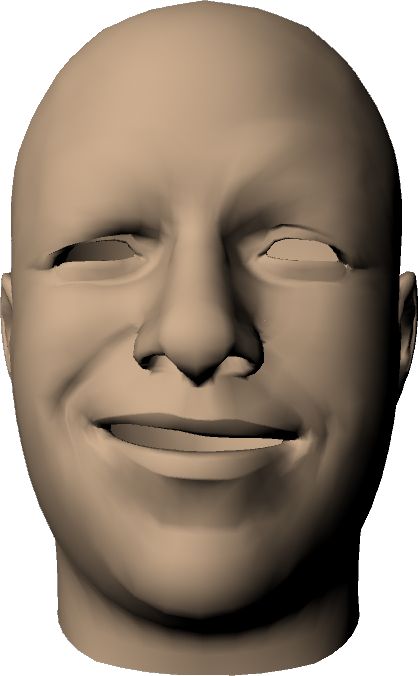}
      \includegraphics[height=0.47\columnwidth]{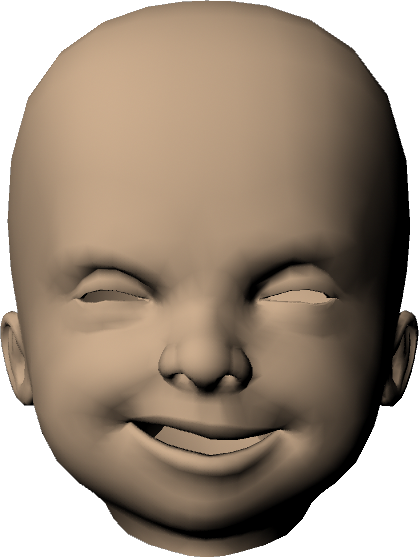}
      \includegraphics[height=0.47\columnwidth]{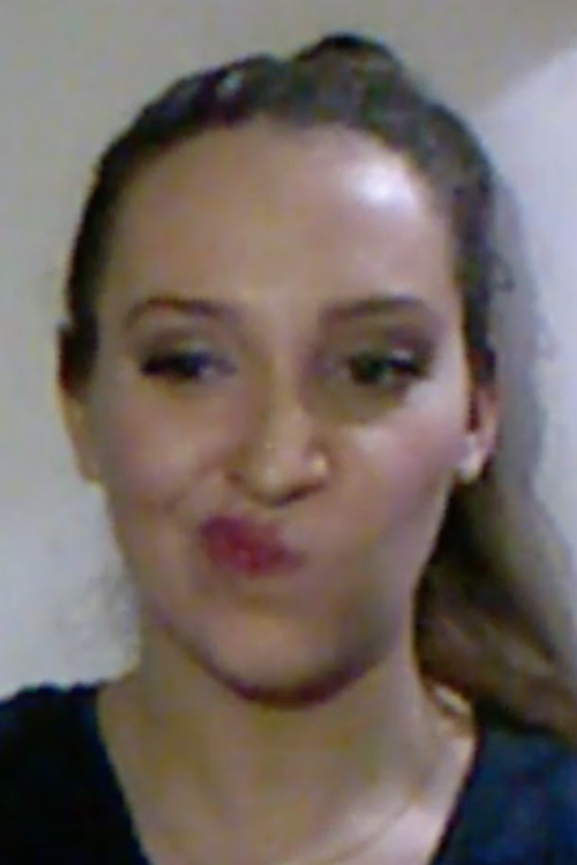}
      \includegraphics[height=0.47\columnwidth]{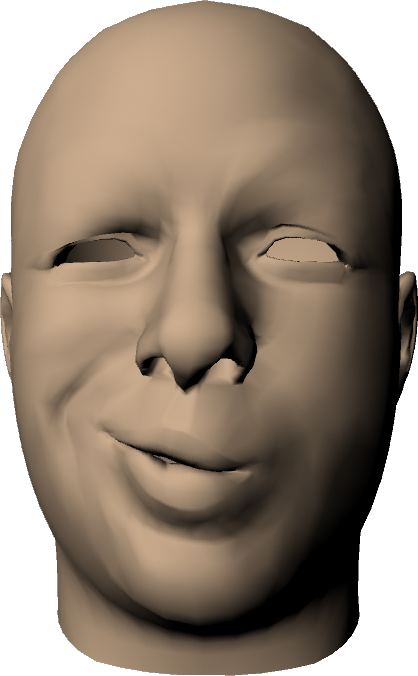}
      \includegraphics[height=0.47\columnwidth]{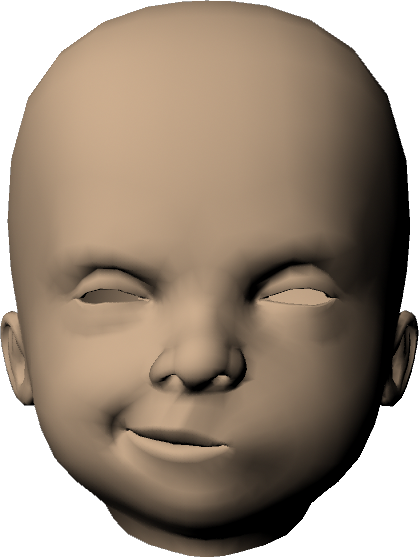}
      \includegraphics[height=0.47\columnwidth]{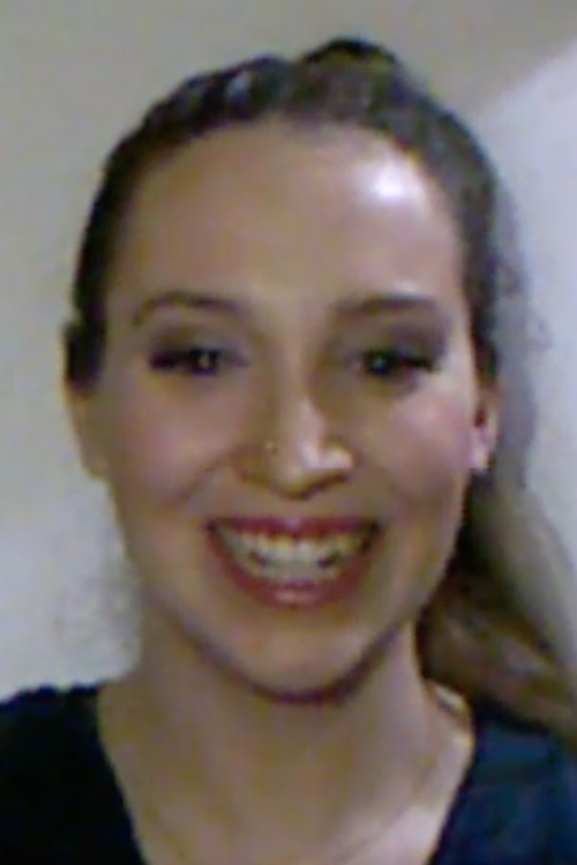}
      \includegraphics[height=0.47\columnwidth]{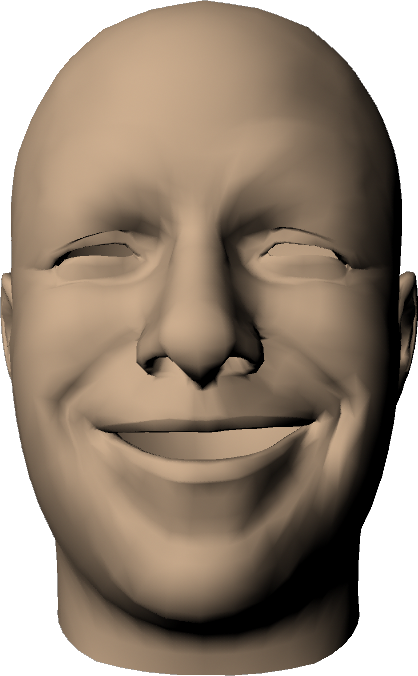}
      \includegraphics[height=0.47\columnwidth]{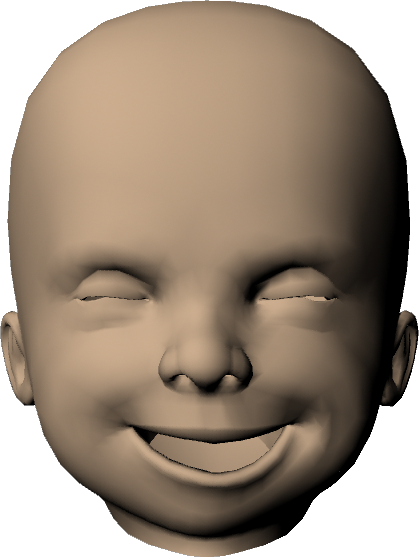}
  \caption{Resulting facial expressions retargeted to different characters.}
   \label{fig3}
\end{figure}

That's it. Based on the aforementioned methodology, the facial animation retargeting method that is presented approximates the output facial expression that appears in a target face model, given a facial expression from a face model with different morphological variations. Example or retargeted motion sequences are shown in Figure \ref{fig3} and in the accompanying video.

\section{Evaluation and Results}
\label{sec4}

\begin{figure}[htb]
  \centering
      \includegraphics[height=0.42\columnwidth]{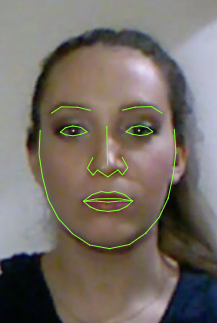}
      \includegraphics[height=0.42\columnwidth]{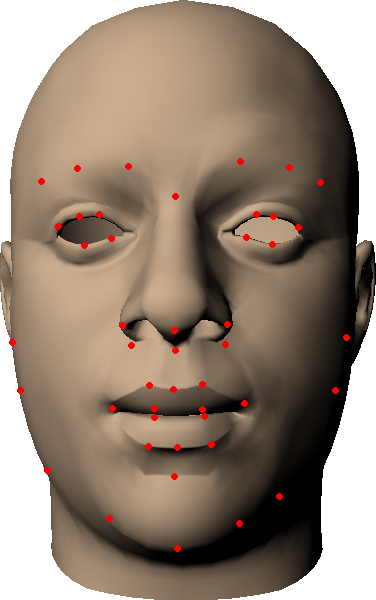}
      \includegraphics[height=0.42\columnwidth]{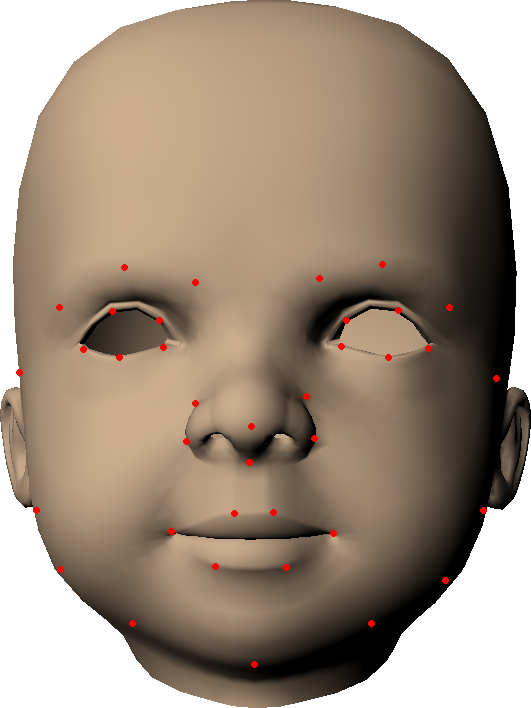}
  \caption{Feature points tracked from (a) an actor, and from the two different models used in this study, (b) man face, and (c) baby face.}
   \label{fig4}
\end{figure}

Two motion sequences were captured for the evaluation process. First, a motion sequence in which an actor performs a monologue (500 frames) was captured. Then, a motion sequence in which the actor performs various facial expressions randomly (1000 frames) was captured. Then, by using a variety of face models that contain a reasonable number of blendshapes, the captured motion sequences were retargeted. Table 1 illustrates the characteristics (vertices and number of blendshapes) of the face models that were used in our study and the number of feature points that were assigned to the face models. The positions of the feature points of each face model that used in this paper are shown in Figure \ref{fig4}. Finally, it should be noted that 46 feature points were retrieved from the actor's face by using \cite{ref46}.

\begin{table}[htb]
\centering
\caption{Characteristics (vertices and blendshapes) of face models and feature points that were used in this paper.}
\label{tab1}
\begin{tabular}{| l | | l | | l | |  p{1.2cm} |}
\hline
\textbf{Model} 		& \textbf{Vertices}	& \textbf{Blendshapes}	& \textbf{Feature Points}\\
\hline
\hline
\textbf{Man}		& 2904			& 48					& 45				\\
\hline
\textbf{Baby}		& 1969			& 44					& 37				\\
\hline
\end{tabular}
\end{table}

\begin{figure}[htb]
  \centering
      \includegraphics[width=0.85\columnwidth]{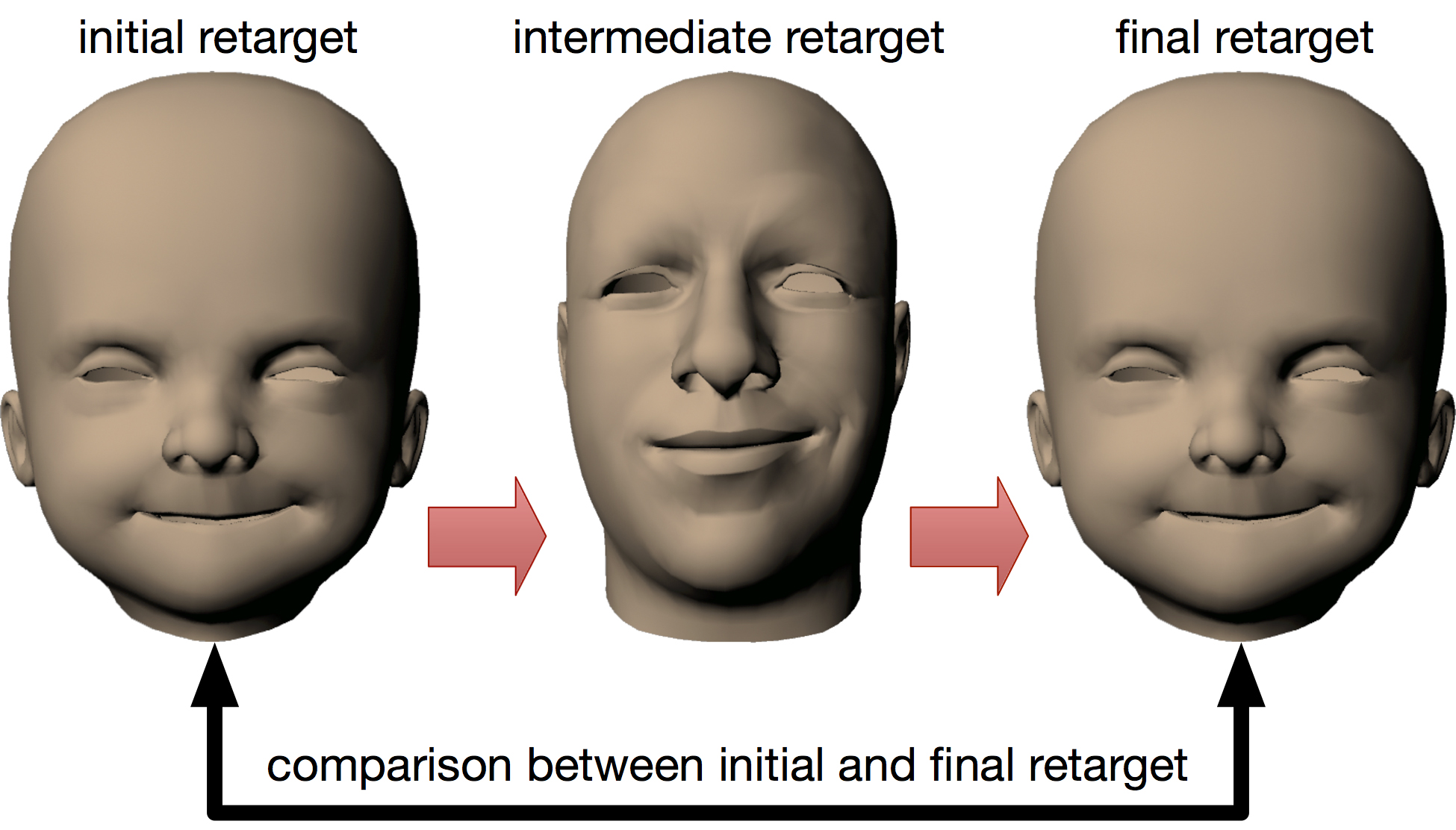}
  \caption{In the evaluation process a motion sequence is retargeted to an intermediate face model and then the intermediate motion is retargeted back to the initial face model.}
   \label{fig5}
\end{figure}

\begin{figure}[htb]
  \centering
      \includegraphics[width= 0.85\columnwidth]{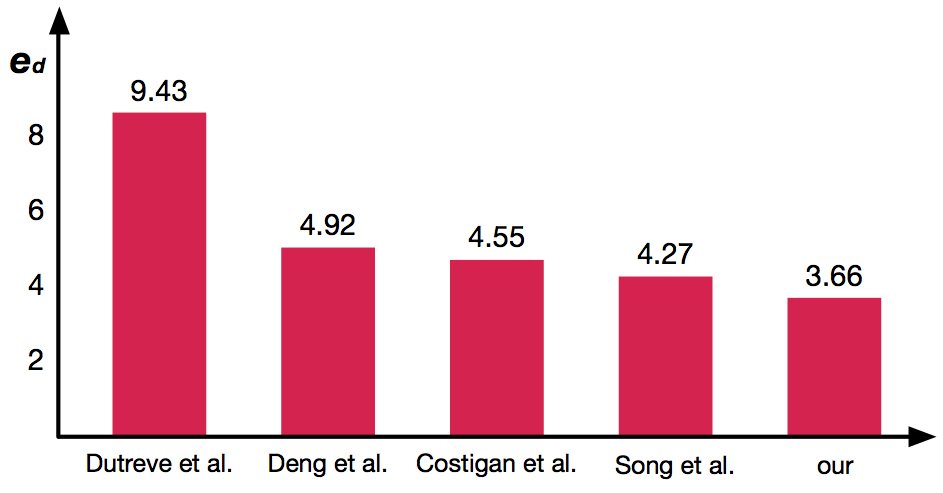}
  \caption{The error $e_d$ between the presented methodology and the methodologies that were proposed by Dutreve et al. \protect\cite{ref18}, Deng et at. \protect\cite{ref19}, Costigan et al. \protect\cite{ref51}, and Song et al. \protect\cite{ref20} based on Equation \protect\ref{eq11}.}
   \label{fig6}
\end{figure}

\begin{figure*}[t]
        \centering
        \begin{subfigure}[b]{0.16\textwidth}
                \includegraphics[width=1\textwidth]{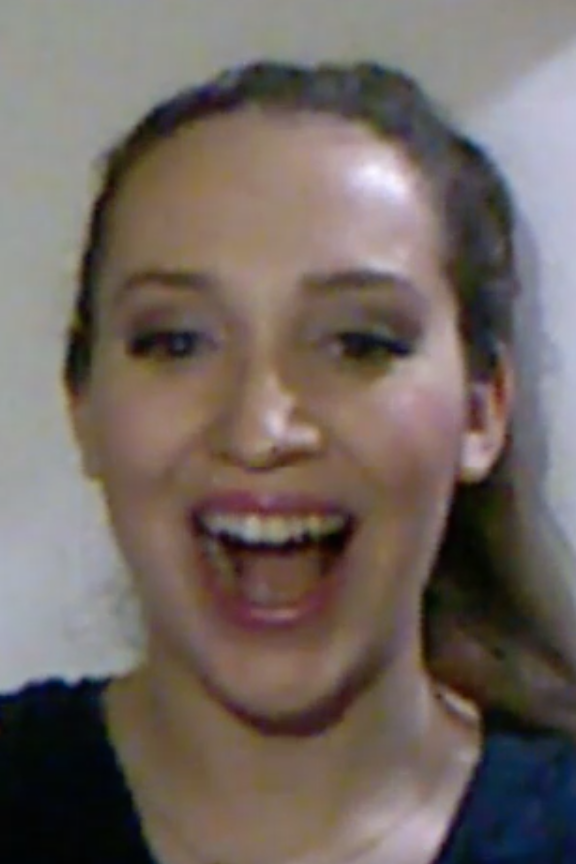}
                \includegraphics[width=1\textwidth]{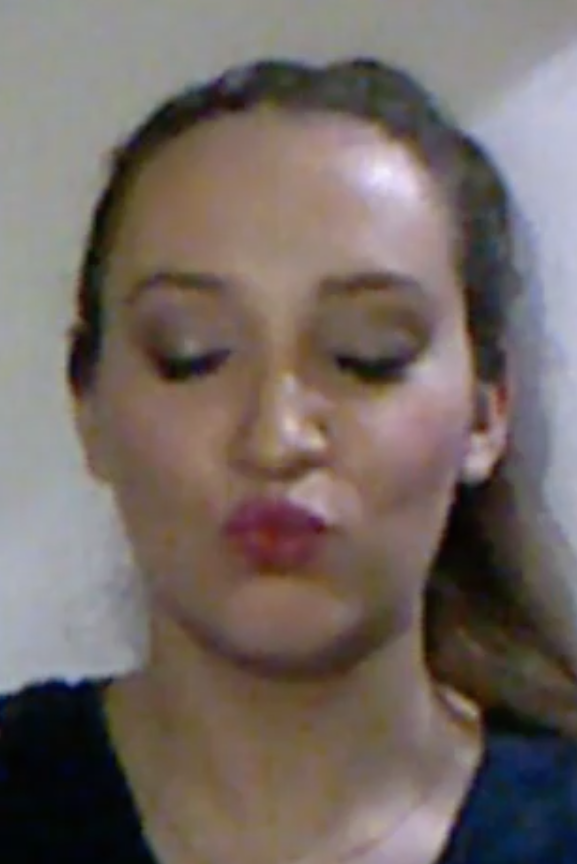}
                \includegraphics[width=1\textwidth]{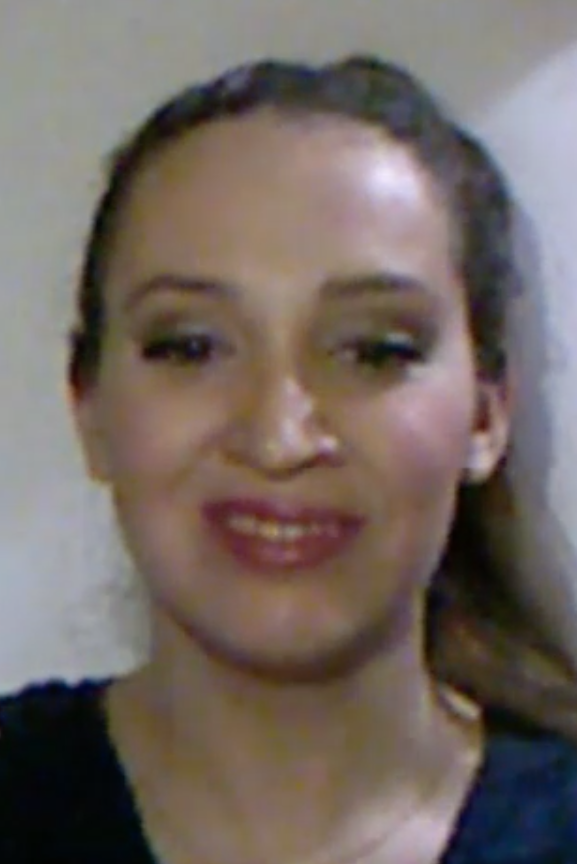}
                \caption{input}
                \label{}
        \end{subfigure}
                \begin{subfigure}[b]{0.16\textwidth}
                \includegraphics[width=0.93\textwidth]{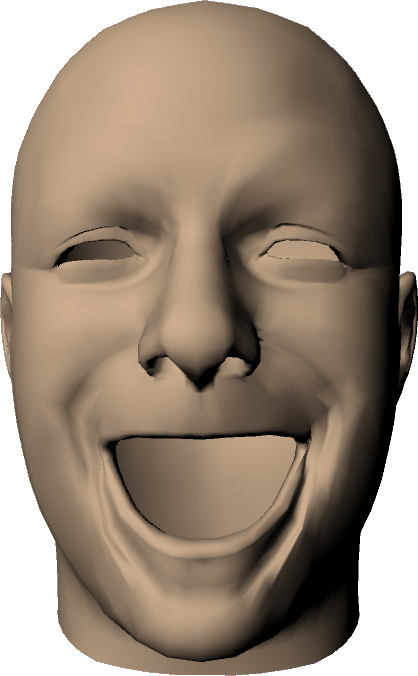}
                \includegraphics[width=0.93\textwidth]{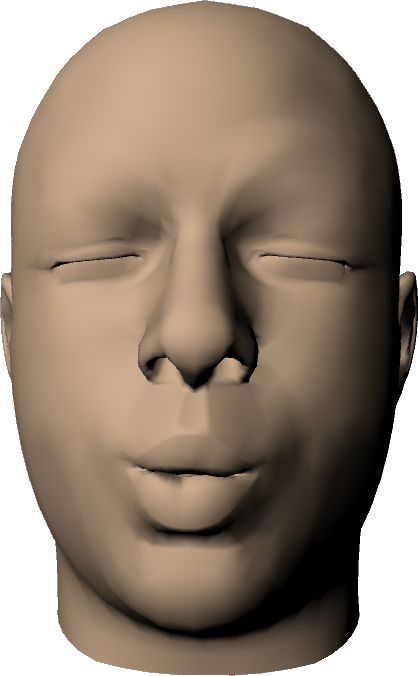}
                \includegraphics[width=0.93\textwidth]{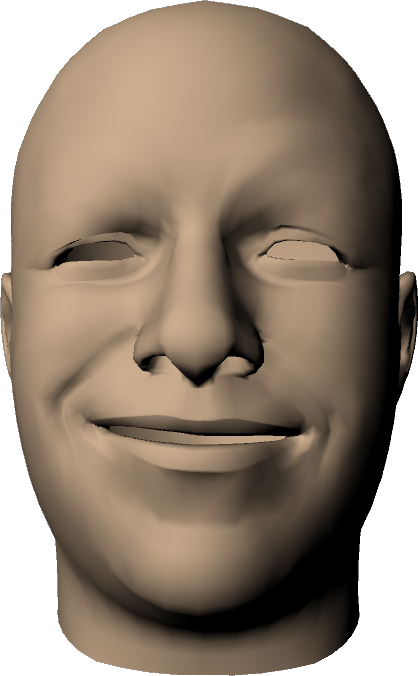}
                \caption{our method}
                \label{}
        \end{subfigure}
        \begin{subfigure}[b]{0.16\textwidth}
                \includegraphics[width=0.93\textwidth]{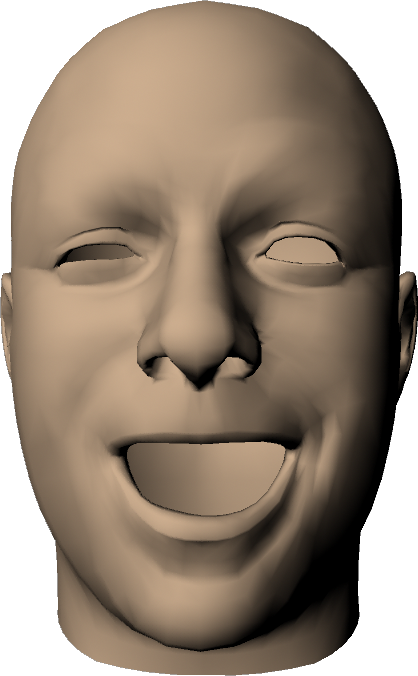}
                \includegraphics[width=0.93\textwidth]{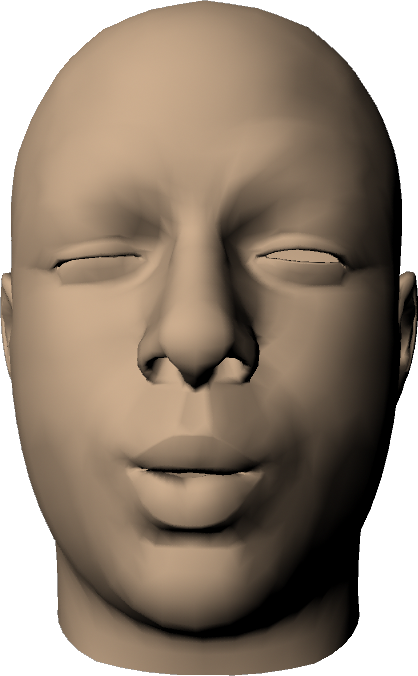}
                \includegraphics[width=0.93\textwidth]{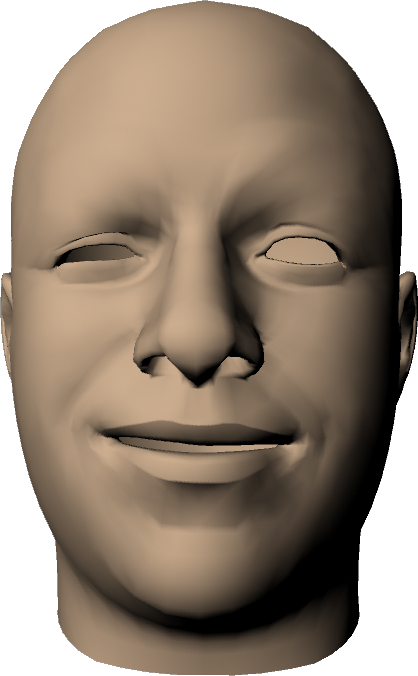}
                \caption{Dutreve et al.}
                \label{}
        \end{subfigure}
        \begin{subfigure}[b]{0.16\textwidth}
                \includegraphics[width=0.93\textwidth]{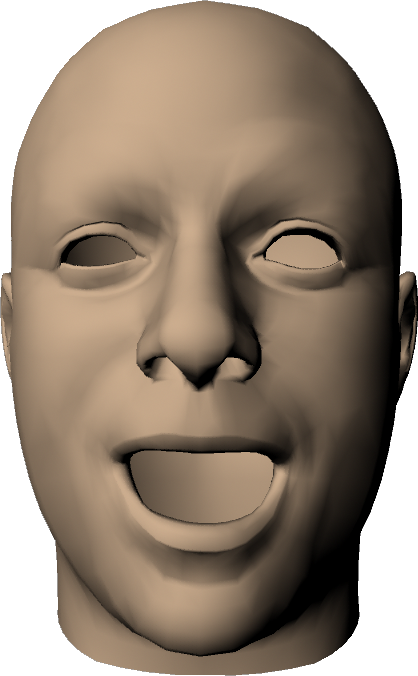}
                \includegraphics[width=0.93\textwidth]{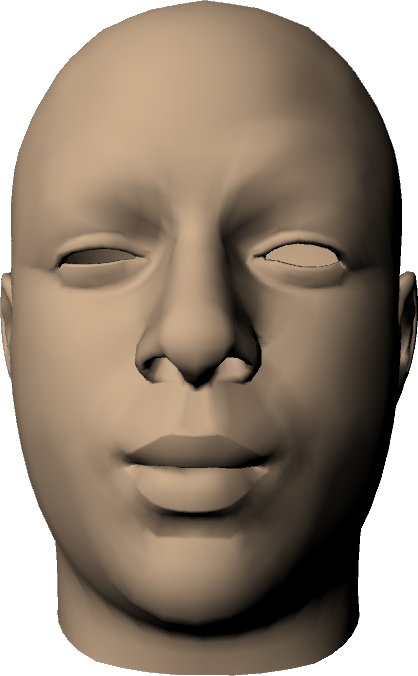}
                \includegraphics[width=0.93\textwidth]{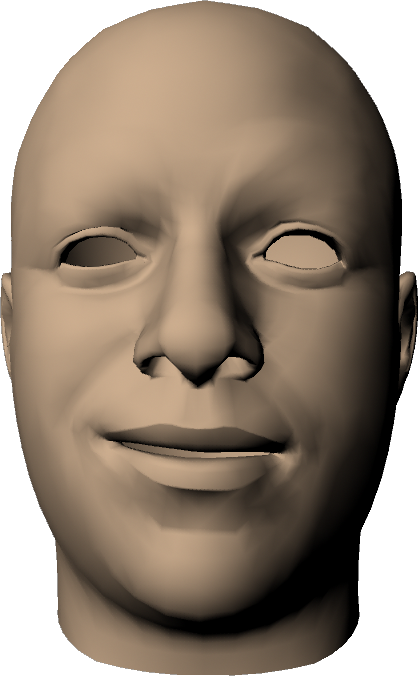}
                \caption{Deng et at.}
                \label{}
        \end{subfigure}
        \caption{Example facial expressions are retargeted to different face models when different methodologies are used. They are, specifically, (a) the input motion, (b) the results of our method, and the results obtained when using the methods that were proposed by (c) \protect\cite{ref18}, and (d) \protect \cite{ref19}.}
        \label{fig7}
\end{figure*}

In order to illustrate the efficiency of the presented facial animation retargeting methodology, an evaluation study was conducted. Specifically, the presented method was evaluated against previously proposed methodologies based on the cyclical process that was presented in \cite{ref13}. For this evaluation process, the presented methodology was compared to \cite{ref18}\cite{ref19}\cite{ref20} and \cite{ref51}. Specifically, the motion is retargeted from an initial human face model to a different (intermediate) target face model. Then, the same procedure was performed inversely. Figure \ref{fig5} illustrates this process. Next, an average vertex displacement error between the initial and the final position of vertices of the two face models was computed. Figure \ref{fig6} illustrates the error that was computed for the presented methodology and the previously proposed methods. Examples that illustrate the initial and final retargeted facial expressions by different methodologies are shown in Figure \ref{fig7} and in the accompanying video. It should be mentioned that for this evaluation process the same number of blendshapes were used for both the initial (baby face) and the intermediate (man face) models. Also, it should be noted that the motion sequence used for this evaluation process was the various random facial expressions (1000 frames). The error between the initial and the final retarget was computed according to the following equation:
\begin{equation}
e_d = \sqrt{\frac{1}{T \times V} \sum_{t=1}^T \sum_{v=1}^V \| p_i^{initial} (t) - p_i^{final} (t) \| ^2 }
\label{eq11}
\end{equation}
where $T$ denotes the total number of frames of a retargeted motion sequence, $V$ denotes the total number of vertices of the face model and, $p_i^{initial}$ and $p_i^{final}$ represent the position of the $i-th$ vertex between the initial and the final motion sequence, respectively.

As the results obtained from the evaluation process (see Figure \ref{fig6}) show, the presented methodology provides closer retargeting to the initial motion than the proposed methods that were previously examined. Specifically, there is a 61\% improvement in similarity compared to \cite{ref18}, a 26\% improvement in similarity compared to \cite{ref19}, a 20\% improvement in similarity compared to \cite{ref51}, and a 14\% improvement in accuracy compared to \cite{ref20}. Based on these results, it should be mentioned that the presented methodology is able to maintain the correspondence between the semantically similar expressions quite effectively. This means that the presented method provides the best and most stable performance for the facial animation retargeting process.

\section{Conclusions and Future Work}
\label{sec5}
This paper introduced a novel facial animation retargeting method. The presented methodology takes advantage of KPLS to build correspondence between examples of semantically similar facial expressions. Based on the presented methodology, the facial animation that is captured from an actor can be retargeted efficiently to a variety of facial models that have different morphological variations.
In our future work, we will continue to work on the facial animation retargeting pipeline. There are various issues that we would like to implement in our current solution. Specifically, we would like to implement a time-warping functionality as introduced in \cite{ref20} in order to transfer effectively certain characteristics of face motion. We assume that such functionality would provide an enhancement to the realism of the motion sequence. The reason is that it could be beneficial for the time varying characteristics that each different face model provides to the final motion. Another issue on which we wish to concentrate is the content retargeting, as introduced in \cite{ref21}. We assume that by incorporating emotional enhancement in our methodology, more realistic facial expressions can be produced. Finally, we would like to incorporate intuitive motion editing techniques in our method, such as those proposed in \cite{ref22}\cite{ref24}. Such techniques could help animators to edit the motion data effectively and easily. We assume that by incorporating all of the aforementioned extensions into our methodology we can provide a powerful tool that could be used in the specific industry.

\bibliographystyle{acmsiggraph}
\nocite{*}
\bibliography{template}

\end{document}